\documentclass[prd,floatfix,preprintnumbers,letterpaper,showpacs]{revtex4}
\usepackage{graphicx}
\usepackage{epsfig}
\usepackage{amsfonts}

\newcommand{\lP}{\ell_{\rm P}}
\newcommand{\md}{{\rm d}}
\newcommand{\tr}{\mathop{\rm tr}}

\begin{document}

\title{Dirac Fields in Loop Quantum Gravity and Big Bang Nucleosynthesis}
\author{Martin Bojowald}
\affiliation{Institute for Gravitation and the Cosmos, The Pennsylvania State University,
104 Davey Lab, University park, PA  ~~16802}
\author{Rupam Das} 
\affiliation{Department of Physics and Astronomy, Vanderbilt University,
Nashville, TN  ~~37235}
\affiliation{Institute for Gravitation and the Cosmos, The Pennsylvania State University,
104 Davey Lab, University park, PA  ~~16802} \author{Robert J. Scherrer} 
\affiliation{Department of Physics and Astronomy, Vanderbilt University,
Nashville, TN  ~~37235}

\begin{abstract}
 Big Bang nucleosynthesis requires a fine balance between equations of
 state for photons and relativistic fermions. Several corrections to
 equation of state parameters arise from classical and quantum
 physics, which are derived here from a canonical perspective. In
 particular, loop quantum gravity allows one to compute quantum
 gravity corrections for Maxwell and Dirac fields. Although the
 classical actions are very different, quantum corrections to the
 equation of state are remarkably similar.  To lowest order, these
 corrections take the form of an overall expansion-dependent
 multiplicative factor in the total density.  We use these results,
 along with the predictions of Big Bang nucleosynthesis, to place
 bounds on these corrections and especially the patch size of discrete
 quantum gravity states.
\end{abstract}


\maketitle
\section{INTRODUCTION}
\label{sec:INTRODUCTION}

Much of cosmology is well-described by a space-time near a spatially
isotropic Friedmann--Robertson--Walker models with line elements
\begin{equation}
 \md s^2= -\md\tau^2+a(\tau)^2\left(\frac{\md
r^2}{1-kr^2}+r^2(\md\vartheta^2+ \sin^2\vartheta\md\varphi^2)\right)
\end{equation}
where $k=0$ or $\pm 1$, sourced by perfect fluids with equations of
state $P=w\rho$. Such an equation of state relates the matter pressure
$P$ to its energy density $\rho$ and captures the thermodynamical
properties in a form relevant for isotropic space-times in general
relativity. Often, one can assume the equation of state parameter $w$
to be constant during successive phases of the universe evolution,
with sharp jumps between different phases such as $w=-1$
during inflation, followed by $w=\frac{1}{3}$ during radiation
domination and $w=1$ during matter domination.

Observationally relevant details can depend on the precise values of
$w$ at a given stage, in particular if one uses an effective value
describing a mixture of different matter components. For instance,
during big bang nucleosynthesis one is in a radiation dominated phase
mainly described by photons and relativistic fermions. Photons,
according to Maxwell theory, have an exact equation of state parameter
$w=\frac{1}{3}$ as a consequence of conformal invariance of the
equations of motion (such that the stress-energy tensor is
trace-free). For fermions the general equation of state is more
complicated and non-linear, but can in relativistic regimes be
approximately given by the same value $w=\frac{1}{3}$ as for photons.
In contrast to the case of Maxwell theory, however, there is no strict
symmetry such as conformal invariance which would prevent $w$ to take
a different value. It is one of the main objectives of the present
paper to discuss possible corrections to this value.

For big bang nucleosynthesis, it turns out, the balance between
fermions and photons is quite sensitive. In fact, different values for
the equation of state parameters might even be preferred
phenomenologically \cite{FermionBoson}. One possible reason for
different equations of state could be different coupling constants of
bosons and fermions to gravity, for which currently no underlying
mechanism is known. In this paper we will explore the possibility
whether quantum gravitational corrections to the equations of state
can produce sufficiently different values for the equation of state
parameters. In fact, since the fields are governed by different
actions, one generally expects different, though small, correction
terms which can be of significance in a delicate balance.  Note that
we are not discussing ordinary quantum corrections of quantum fields
on a classical background. Those are expected to be similar for
fermions and radiation in relativistic regimes. We rather deal with
quantum gravity corrections in the coupling of the fields to the
space-time metric, about which much less is known a priori. Thus,
different proposals of quantum gravity may differ at this stage,
providing possible tests.

An approach where quantum gravitational corrections can be computed is
loop quantum cosmology \cite{LivRev}, which specializes loop quantum
gravity \cite{Rov,ALRev,ThomasRev} to cosmological regimes. In such a
canonical quantization of gravity, equations of state must be computed
from matter Hamiltonians rather than covariant stress-energy tensors.
Quantum corrections to the underlying Hamiltonian then imply
corrections in the equation of state. This program was carried out for
the Maxwell Hamiltonian in \cite{MaxwellEOS}, and is done here for
Dirac fermions. There are several differences between the treatment of
fermions and other fields, which from the gravitational point of view
are mainly related to the fact that fermions, in a first order
formulation, also couple to torsion and not just the curvature of
space-time. After describing the classical derivation of equations of
state as well as steps of a loop quantization and its correction
terms, we use big bang nucleosynthesis constraints to see how
sensitively we can bound quantum gravity parameters.

We have aimed to make the paper nearly self-contained and included
some of the technical details. Secs.~\ref{sec:canonical Formulation}
on the canonical formulation of fermions, \ref{sec:MODIFICATIONS} on
quantum corrections from loop quantum gravity and \ref{sec:BBN} on the
analysis of big bang nucleosynthesis can, however, be read largely
independently of each other by readers only interested in some of the
aspects covered here.  We will start with general remarks on the
physics underlying the problem.

\section{The physical setting}
\label{sec:physics}

Big bang nucleosynthesis happens at energy scales $E_{\rm BBN}\sim
{\rm MeV}$ which are large, but still tiny compared to the Planck
energy $M_{\rm P}$. Also the universe has already grown large compared
to the Planck length $\ell_{\rm P}$ at this stage, and space-time
curvature is small. One may thus question why quantum gravity should
play any role. There is certainly a fine balance required for
successful big bang nucleosynthesis, but the expected quantum gravity
terms of the order $E/M_{\rm P}$, obtained based on dimensional
arguments, would have no effect.

However, dimensional arguments do not always work, in particular if
more than two parameters $L_I$ of the same dimension, or any large
dimensionless numbers are involved. Then, precise calculations have to
be done to determine which geometric means $\prod_IL_I^{x_I}$ with
$\sum_Ix_I=0$ may appear as coefficients, or which powers of
dimensionless numbers occur as factors of correction terms. In loop
quantum gravity, we are in such a situation: there is the macroscopic
length scale $L$, which in our case we can take as the typical wave
length of fields during nucleosynthesis, and also the Planck length
$\ell_{\rm P}=\sqrt{G\hbar}$ which arises due to the presence of
Newton's constant $G$ and Planck's constant $\hbar$. In addition,
there is a third and in general independent scale $\ell$ given by the
microscopic size of elementary spatial patches in a quantum gravity
state. This is a new feature of the fundamentally discrete theory, for
which the precise state of quantum gravity plays an important
role. Although $\ell$ must be proportional to the Planck length, its
specific value for a given state can differ numerically. Then, a
detailed calculation must show how $L$, $\ell$ and $\ell_{\rm P}$
appear in quantum gravity corrections and which numerical values may
arise.

Alternatively, one can work with only two length scales, $L$ and
$\ell_{\rm P}$, but one has to deal with a large dimensionless parameter
${\cal N}$ given by the number of discrete patches of the underlying
state in the volume considered, for instance a volume of the size
$L^3$ such that ${\cal N}=L^3/\ell^3$. Examples of cosmological
phenomena are known where this does play a role for quantum gravity
corrections \cite{InhomEvolve,SILAFAE}, and here we analyze which
features arise in the presence of fermions and especially for big bang
nucleosynthesis.

There are precedents where such considerations have played important
roles. Best known is the evidence for the atomic nature of matter
derived by Einstein from the phenomenon of Brownian motion. Also here,
there are several orders of magnitude between the expected size of
molecules and the resolution of microscopes at that time. However,
there is also a large number of molecules which by their sheer number
can and do leave sizeable effects on much larger suspended
particles. There is, of course, never a guarantee that something
analogous has to happen elsewhere. But this is to be checked by
calculations and cannot always be ruled out based only on dimensional
arguments.

We can use the result of corrections to the equation of state of
radiation derived in \cite{MaxwellEOS}, but a new analysis is required
for fermions and their specific action. Even in relativistic regimes,
the coupling of fermions to gravity differs from other fields, e.g.\
by torsion contributions which arise already from the kinetic term of
the Dirac action. One could thus expect that quantum corrections for
fermions differ from those to radiation and thus, by throwing off the
balance during nucleosynthesis, possibly enhance the effect of quantum
gravity corrections. Whether or not this happens cannot be decided
without detailed calculations as they are reported and applied here.

\section{Canonical Formulation of Dirac Fermions}
\label{sec:canonical Formulation}

For fermions, one has to use a tetrad $e^I_{\mu}$ rather than a
space-time metric $g_{\mu\nu}$, which are related by
$e^I_{\mu}e^I_{\nu}=g_{\mu\nu}$, in order to formulate an action with
the appropriate covariant derivative of fermions. This naturally leads
one to a first-order formalism of gravity in which the basic
configuration variables are a connection 1-form and the tetrad. In
vacuum the connection would, as a consequence of field equations, be
the torsion-free spin connection compatible with the tetrad.  In the
presence of matter fields which couple directly to the connection,
such as fermions, this is no longer the case and there is torsion
\cite{SpinTorsion, FermionHolst, FermionAshtekar}. 

\subsection{Non-minimally Coupled Einstein-Dirac Action}

A canonical analysis of gravity minimally coupled to fermions is
presented in detail in \cite{FermionHolst}. In \cite{FermionAshtekar},
a non-minimally coupled action is proposed in order to eliminate
violations of parity. Here, we first present a general canonical
analysis providing a consistent canonical formulation for all coupling
parameters. At the end of this section, we briefly comment on parity
which plays a role in this context.

The basic configuration variables in a Lagrangian formulation of
fermionic field theory are the Dirac bi-spinor $\Psi = \left(\psi \ \
\eta\right)^T$ and its adjoint
$\overline{\Psi}=\left(\Psi^{*}\right)^{T}\gamma^{0}$ with
$\gamma^{\alpha}$ being the Minkowski Dirac matrices. The 2-component
${\rm SL}(2,{\mathbb C})$-spinors $\psi$ and $\eta$ will be used later
on in a Hamiltonian decomposition of the action. Being interested in
applications to highly relativistic regimes, we only deal with
massless fermions. Their non-minimal coupling to gravity can then be
expressed, in the notation of \cite{FermionAshtekar}, by the total
action
\begin{eqnarray}
\label{nonminimalaction}
S\left[e,\omega,\Psi\right]&=& S_{\rm G}\left[e,\omega\right]+
S_{\rm F}\left[e,\omega,\Psi\right] \nonumber\ \\ &=& 
\frac{1}{16\pi G} \int_{M}\md^{4}x \;ee^{\mu}_{I}e^{\nu}_{J}
P^{IJ}_{\ \ \ KL}F^{\ \ KL}_{\mu
\nu}(\omega) \nonumber\ \\ &&+ \frac{1}{2}i\int_{M} \md^{4}x \;
e\left[\overline{\Psi}\gamma^{I}e^{\mu}_{I}\left(1-\frac{i}{\alpha}
\gamma_{5}\right)\nabla_{\mu}\Psi -
\overline{\nabla_{\mu}\Psi}\left(1-\frac{i}{\alpha}\gamma_{5}\right)
\gamma^{I}e^{\mu}_{I}\Psi\right],
\end{eqnarray}
where $\alpha$ is the parameter for non-minimal coupling and minimal
coupling is reproduced for $\alpha\to\infty$. The action is composed
of the matter contribution $S_{\rm F}$ resulting from the fermion
field and the gravitational contribution $S_{\rm G}$ expressed in
terms of
\begin{equation}
\label{PIJ}
P^{IJ}_{\ \ \ KL}=\delta^{[I}_{K} \delta^{J]}_{L} - \frac{1}{\gamma}
\frac{\epsilon^{IJ}_{\ \ KL}}{2}\quad,\ \ \ {P^{-1}_{\ \ \
IJ}}^{KL}=\frac{\gamma^{2}}{\gamma^{2}+1}\left(\delta^{[K}_{I}
\delta^{L]}_{J} + \frac{1}{\gamma} \frac{\epsilon_{IJ}^{\ \ \
KL}}{2}\right)
\end{equation}
where $\gamma$ is the Barbero--Immirzi parameter
\cite{Immirzi,AshVarReell}.

The specific form
of the gravitational action is the one given by Holst
\cite{HolstAction}, formulated in terms of a tetrad field
$e^{\mu}_{I}$ with inverse $e^{I}_{\mu}$, whose determinant is $e$.
(Thus, the space-time metric is $g_{\mu\nu}=e^I_{\mu}e^I_{\nu}$.  For
all space-time fields, $I,J,\ldots=0,1,2,3$ denote internal Lorentz
indices and $\mu, \nu, \ldots=0,1,2,3$ space-time indices.)  The
Lorentz connection $\omega_{\mu}^{IJ}$ in this formulation is an
additional field independent of the triad, and $F^{KL}_{\mu
  \nu}(\omega)= 2\partial_{[\mu}\omega^{IJ}_{\nu]}+
\left[\omega_{\mu},\omega_{\nu}\right]^{IJ}$ is its curvature.  It
also determines the covariant derivative $\nabla_{\mu}$ of Dirac
spinors by
\begin{equation}
\label{covariantderivative1}
 \nabla_{\mu}\equiv \partial_{\mu} + \frac{1}{4}\omega^{IJ}_{\mu}
 \gamma_{[I} \gamma_{J]}\quad, \ \ \ \ \
 \left[\nabla_{\mu},\nabla_{\nu}\right] = \frac{1}{4}F^{IJ}_{\mu
 \nu}\gamma_{[I} \gamma_{J]}
\end{equation}
in terms of Dirac matrices $\gamma_I$ (which will always carry an
index such that no confusion with the Barbero--Immirzi parameter
$\gamma$ can arise).

\subsection{Dirac Hamiltonian}

As usually, a Hamiltonian formalism of gravity requires a space-time
foliation $\Sigma_t\colon t={\rm const}$ such that one can introduce
fields and their rates of change, which will provide canonical
variables. This is done by referring to a time function $t$ as well as
a time evolution vector field $t^{\mu}$ such that
$t^{\mu}\nabla_{\mu}t=1$. Rates of change of spatial fields will then be
associated with their derivatives along $t^a$.

For convenience, one decomposes $t^{\mu}$ into normal and tangential
parts with respect to $\Sigma_{t}$ by defining the lapse function $N$
and the shift vector $N^{a}$ such that $t^{\mu}= Nn^{\mu}+ N^{\mu}$
with $N^{\mu}n_{\mu}=0$. Here, $n^{\mu}$ is the unit normal vector
field to the hypersurfaces $\Sigma_{t}$. The space-time metric
$g_{\mu\nu}$ induces a spatial metric $q_{\mu\nu}(t)$ on $\Sigma_t$ by
the formula $g_{\mu\nu}=q_{\mu\nu}-n_{\mu}n_{\nu}$. This is one of the
basic fields of a canonical formulation, and its momentum will be
related to $\dot{q}_{ab}$ defined as the Lie derivative of $q_{ab}$
along $t^a$.  Since contractions of $q_{\mu\nu}$ and $N^{\mu}$ with
the normal $n^{\mu}$ vanish, they give rise to spatial tensors
$q_{ab}$ and $N^a$.

In our case, we are using a tetrad formulation, where $e^I_{\mu}$
provides a map from the tangent space of space-time to an internal
Minkowski space. The space-time foliation thus requires an associated
space-time splitting of the Minkowski space. This takes the form of a
partial gauge fixing on the internal vector fields of the tetrad: the
directions (or rather boosts) of tetrad fields can no longer be chosen
arbitrarily. Instead, we decompose the tetrad into a fixed internal
unit time-like vector field and a triad on the space $\Sigma_t$.  We
choose the internal vector field to be constant, $n_{I}=-\delta_{I,0}$
with $n^{I}n_{I}=-1$. Then, we allow only those tetrads which are
compatible with the fixed $n^{I}$ in the sense that $n^{a}=
n^{I}e_{I}^{a}$ is the unit normal to the given foliation.  This
implies that $e^{a}_{I}= {\cal E}^{a}_{I} - n^{a}n_{I}$ with ${\cal
E}^{a}_{I}n_{a}={\cal E}^{a}_{I}n^{I}=0$ so that ${\cal E}^{a}_{I}$ is
a spatial triad.
 
Now, using $e^{a}_{I}= {\cal E}^{a}_{I} - n^{a}n_{I}$ with
$n_{I}=-\delta_{I,0}$ and $n^{a}=N^{-1}(t^{a}-N^{a})$ we can decompose
the non-minimally coupled Dirac action and write it in terms of
spatial fields only:
\begin{equation}
\label{action2}
S_{\rm Dirac}= \frac{i}{2}\int_{M} \md^{4}x \;
N\sqrt{q}\left(\overline{\Psi}\gamma^{I}({\cal E}^{a}_{I}+N^{-1}(t^{a}-N^{a}))
\left(1-\frac{i}{\alpha}\gamma_{5}\right)\nabla_{a}\Psi -
\overline{\nabla_{a}\Psi}\left(1-\frac{i}{\alpha}\gamma_{5}\right)
\gamma^{I}({\cal E}^{a}_{I}+N^{-1}(t^{a}-N^{a}))\Psi\right),\
\end{equation}
where the space-time
determinant is factorized as $e=N\sqrt{q}$ with the determinant $q$ of
the spatial metric.
These terms can be decomposed into several terms containing the ${\rm
  SL}(2,{\mathbb C})$-spinors $\psi$ and $\eta$ instead of the Dirac
spinor $\Psi$:
\begin{eqnarray}
\label{totalpsiexpansion}
\frac{i}{2}N\sqrt{q}n^{a}(\overline{\Psi}\gamma^{0}\nabla_{a}\Psi- c.c.)+ \frac{i}{2}N{\cal E}^{a}_{i}(\overline{\Psi}\gamma^{i}\nabla_{a}\Psi-c.c.)
=\sqrt{q}\left(-{\textstyle\frac{1}{2}}i(\psi^{\dagger}\dot{\psi}+
\eta^{\dagger}\dot{\eta}- c.c.)+
{\textstyle\frac{1}{4}}\epsilon_{mnk}\omega_{t}^{\ mn}J^{k}\right) \nonumber\ \\ 
-\sqrt{q}N^{a}\left(-{\textstyle\frac{1}{2}}i(\psi^{\dagger}\partial_{a}{\psi}+ 
\eta^{\dagger}\partial_{a}{\eta}- c.c.)  
+{\textstyle\frac{1}{4}}\epsilon_{mnk}\omega_{a}^{\ mn}J^{k}\right)-
N\sqrt{q}{\cal E}^{a}_{i}\left(-{\textstyle\frac{1}{2}}i
(-\psi^{\dagger}\sigma^{i}\partial_{a}\psi+
\eta^{\dagger}\sigma^{i}\partial_{a}\eta-c.c.)\right. \nonumber\ \\ 
-\left({\textstyle\frac{1}{4}}\epsilon^{i}_{\ mn}\omega_{a}^{\ mn}
(\psi^{\dagger}\psi-\eta^{\dagger}\eta)
+{\textstyle\frac{1}{2}}\epsilon^{i}_{\ mn}\omega_{a}^{\ m0}J^{n}\right) \,,
\end{eqnarray} 
and similarly an expansion of the terms involving the non-minimal
coupling parameter $\alpha$ in the action (\ref{action2}) gives
\begin{widetext}
\begin{eqnarray}
\label{nonminimalactionexpansion}
&&\frac{1}{2\alpha}\sqrt{q}((t^{a}-N^{a})\left(\overline{\Psi}
\gamma^{0}\gamma_{5}\nabla_{a}\Psi-\overline{\nabla_{a}\Psi}
\gamma_{5}\gamma^{0}\Psi \right)
+N{\cal E}^{a}_{i}\left(\overline{\Psi}\gamma^{i}\gamma_{5}\nabla_{a}
\Psi-\overline{\nabla_{a}\Psi}\gamma_{5}\gamma^{i}\Psi \right))\nonumber\ \\ 
&=& \frac{1}{2\alpha}\sqrt{q}(t^{a}-N^{a})(\psi^{\dagger}\partial_{a}\psi-
\eta^{\dagger}\partial_{a}\eta+(\partial_{a}\psi^{\dagger})\psi - 
(\partial_{a}\eta^{\dagger})\eta-\omega_{a}^{\ j0}J^{j})\nonumber\ \\
&&+\frac{N}{2\alpha}\sqrt{q}{\cal E}^{a}_{i}\left(\psi^{\dagger}
\sigma^{i}\partial_{a}\psi+\eta^{\dagger}\sigma^{i}\partial_{a}\eta+
(\partial_{a}\psi^{\dagger})\sigma^{i}\psi+(\partial_{a}\eta^{\dagger})
\sigma^{i}\eta+\omega_{a}^{\ ik}J_{k}-\omega_{a}^{\ i0}J_{0}\right)\,,
\end{eqnarray}
\end{widetext}
where $c.c.$ denotes complex conjugation and we have introduced the
fermion current
$J^{k}:=\psi^{\dagger}\sigma^{k}\psi+\eta^{\dagger}\sigma^{k}\eta$ and
the time component $J_{0} := \psi^{\dagger}\psi-\eta^{\dagger}\eta $.
Details of this calculation as well as the corresponding canonical
decompositions of the gravitational part can be found in
\cite{FermionHolst}.

We also refer to \cite{FermionHolst} for the definition of canonical
gravitational variables as they appear also in the matter action.  In
terms of the connection components $\omega_{\mu}^{IJ}$ in
Eq.~(\ref{covariantderivative1}), we define
\begin{eqnarray}
\label{definitionsofgammaandk1}
\Gamma^i_b:=-\frac{1}{2}\epsilon^i{}_{jk}\omega_{b}^{jk}
\quad,\quad
K_b^i:=-\omega_{b}^{\ i0}
\end{eqnarray} 
and the Ashtekar--Barbero connection \cite{AshVar,AshVarReell}
\begin{eqnarray}
\label{abconnection}
A^{i}_{a} := \Gamma^{i}_{a} + \gamma K^{i}_{a} = 
-\frac{1}{2}\epsilon^{i}_{\ kl}\omega_{a}^{\ kl}-\gamma
\omega_{a}^{\ i0}
\end{eqnarray}
with the Barbero--Immirzi parameter $\gamma$ \cite{Immirzi}.
This connection is important because it provides a convenient
canonical structure with Poisson brackets
\begin{equation} \label{Poisson}
 \{A_a^i(x),E^b_j(y)\} = 8\pi \gamma G \delta^b_a\delta^i_j\delta(x-y)
\end{equation}
with the densitized triad $E^b_j:=\sqrt{q} {\cal E}^b_j$.
Moreover, following the steps presented in \cite{FermionHolst}, it is
straightforward to show that the torsion contribution to the spin
connection $\Gamma$ is given by
\begin{eqnarray}
\label{cbk5}
C^{j}_{a} := \frac{\gamma
  \kappa}{4(1+\gamma^{2})}\left(\left(1-\frac{\gamma}{\alpha}\right) \ 
\epsilon^{j}_{\
    kl}e_{a}^{k}J^{l}-\left(\gamma+\frac{1}{\alpha}\right)e_{a}^{j} 
J^{0}\right)\,.
\end{eqnarray}
One can then show that
$\Gamma_{a}^{i}=\widetilde{\Gamma}_{a}^{i}+C_{a}^{i}$, with
$\widetilde{\Gamma}_{a}^{i}$ being the torsion-free spin connection
compatible with the co-triad $e_{a}^{i}$.

For a fixed value of $\gamma$, the above equation for torsion reduces
to that for minimal coupling as $\alpha \rightarrow \infty$. As noted
in \cite{FermionAshtekar}, based on a Lagrangian analysis, another
interesting case is $\alpha=\gamma$, which reduces $C^{j}_{a}$ to
$-\frac{\kappa}{4}e_{a}^{i}J^{0}$, making torsion
$\gamma$-independent. The non-minimal coupling is important for the
behavior of the action under parity reversal, and it is the value
$\alpha=\gamma$ which provides parity invariance of the combined
system of gravity and fermions. This may seem surprising because
neither vacuum gravity nor the minimal Dirac action on a fixed
background violate parity, and thus the extra term of non-minimal
coupling seems to introduce parity violation. As
\cite{FermionAshtekar} in the Lagrangian picture and
\cite{FermionHolst} in the Hamiltonian picture show, however, the
torsion introduced by coupling fermions to gravity also introduces
parity violation in the Holst action, which has to be canceled by the
additional term of non-minimal coupling. (The action with minimal
coupling was called ``not fully consistent'' in
\cite{FermionAshtekar}, while our analysis is consistent for any value
of $\alpha$. To avoid potential confusion, one should first note that
\cite{FermionAshtekar} starts from the Einstein--Cartan action with
minimal coupling, not directly from a Holst action as here. Parity
properties are different in both cases, and the observation of
\cite{FermionAshtekar} refers only to this behavior. The Holst action
with minimal coupling would indeed be inconsistent with parity
preservation, but it does not present an inconsistency of the overall
framework unless one explicitly requires parity preservation. In fact,
parity violation is expected at the onset of BBN due to weak
interactions. Anyway, the following discussion of our paper can remain
unchanged for different values of $\alpha$ and thus applies directly
to the parity preserving case.)

Upon inserting (\ref{cbk5}) in (\ref{totalpsiexpansion}) and
(\ref{nonminimalactionexpansion}), the action (\ref{action2}) takes
the form
\begin{eqnarray}
\label{splitaction}
S_{\rm Dirac}\left(P^{a}_{i},\Gamma^{i}_{a},\psi, {\psi^{\dagger}}, \eta, 
{\eta^{\dagger}}\right) 
&=&-\int \md t \int_{\Sigma_{t}} \md^{3}x \sqrt{q}\left(
\left(i\left(\theta_{L}(\psi^{\dagger}\dot{\psi} 
- \dot{\eta}^{\dagger}{\eta}) - \theta_{R}(\dot{\psi}^{\dagger}{\psi} 
- {\eta}^{\dagger}\dot{\eta}) \right)
-{\textstyle\frac{1}{4}}\epsilon_{mnk}\omega_{t}^{\ mn}J^{k} 
+\frac{1}{2\alpha}\omega_{t}^{\ j0}J_{i}\right)\right.\nonumber\ \\
&& - N^{a}\left(i\left(\theta_{L}(\psi^{\dagger}{\cal D}_{a}\psi-
\overline{{\cal D}_{a}\eta}\eta)+\theta_{R}(\eta^{\dagger}{\cal D}_{a}\eta-
\overline{{\cal D}_{a}\psi}\psi)\right)
-\frac{1}{2}\left(\gamma+\frac{1}{\alpha}\right)K^{i}_{a}\sqrt{q}J_{i}
\right)\nonumber\ \\
&& +N\left(
E^{a}_{i}\left(i\theta_{L}(\psi^{\dagger}\sigma^{i}{\cal
  D}_{a}\psi+\overline{{\cal D}_{a}\eta}\sigma^{i}\eta)-i\theta_{R}
(\eta^{\dagger}\sigma^{i}{\cal
  D}_{a}\eta+\overline{{\cal D}_{a}\psi}\sigma^{i}\psi)\right)-
\frac{1}{2}\left(\gamma+\frac{1}{\alpha}\right) E^{a}_{i}K_{a}^{i}J^{0}
\right.
\nonumber\ \\ 
&& +\left.\left.\frac{1}{2} \epsilon_{lkn}K_{a}^{l}E^{a}_{k}\left(1-
\frac{\gamma}{\alpha}\right)J^{n}\right)\right)\,,
\end{eqnarray}
where $\theta_{L}:=\frac{1}{2}(1+\frac{i}{\alpha})$,
$\theta_{R}:=\frac{1}{2}(1-\frac{i}{\alpha})$ and we have used the
covariant derivatives, ${\cal D}_{a}=\partial_{a}+A^{l}_{a}\tau_{l}$,
related to the Ashtekar-Barbero connection, and
$\tau_{l}=-\frac{1}{2}i\sigma_{l}$.  The first part of the action in
this form shows that momenta of the fermion field are
$\pi_{\xi}=-i\xi^{\dagger}$ of $\xi=\sqrt[4]{q}\psi$ and
$\pi_{\chi}=-i\chi^{\dagger}$ of $\chi=\sqrt[4]{q}\eta$. Here, we are
using half-densitized spinor fields to avoid complex-valued canonical
variables \cite{FermionHiggs}.

Moreover, the action provides the Hamiltonian $H=\int\md^3x N(x)
\delta S/\delta N$.  Together with fermion dependent terms resulting
from the gravitational action, this provides derivative terms and
self-interaction terms in the Hamiltonian
\begin{eqnarray}
\label{dirachamiltonianconstraint3}
 H_{\rm Dirac}
&=& \int_{\Sigma_{t}} \md^{3}xN\left(-\frac{\beta
   E^{a}_{i}}{\sqrt{q}}{\cal
   D}_{a}\left(\pi^{T}_{\xi}\tau^{i}\xi+\pi^{T}_{\chi}
\tau^{i}\chi\right)-i\frac{2 E_{a}^{i}}{\sqrt{q}} 
\left(\theta_{L}\pi^{T}_{\xi}\tau^{i}{\cal D}_{a}\xi 
 -\theta_{R}\pi^{T}_{\chi}\tau^{i}{\cal D}_{a}\chi -
 c.c.\right)\right. \\ 
&&+\left.\frac{\gamma \kappa \beta}{2\sqrt{q}(1+\gamma^{2})}\left(3-
\frac{\gamma}{\alpha}+
2\gamma^{2}\right)(\pi^{T}_{\xi}\tau_{l}\xi+\pi^{T}_{\chi}
\tau_{l}\chi)(\pi^{T}_{\xi}\tau^{l}\xi+\pi^{T}_{\chi}
\tau^{l}\chi)+\frac{3\gamma\kappa}{8\alpha \sqrt{q}}
\left(1-\frac{\gamma}{\alpha}\right)
(\pi^{T}_{\xi}\xi-\pi^{T}_{\chi}\chi)(
\pi^{T}_{\xi}\xi-\pi^{T}_{\chi}\chi)\right)\nonumber 
\end{eqnarray}
with $\beta := \gamma+\frac{1}{\alpha}$.
The top line of this expression is the most important one because its
derivative terms are dominant in relativistic regimes. In addition to
those, we highlight the presence of four-fermion interactions in the
second line, which we summarize as
\begin{equation}
\label{B}
B :=\frac{\gamma \kappa \beta}{2(1+\gamma^{2})}\left(3-
\frac{\gamma}{\alpha}+
2\gamma^{2}\right)(\pi^{T}_{\xi}\tau_{l}\xi+\pi^{T}_{\chi}
\tau_{l}\chi)(\pi^{T}_{\xi}\tau^{l}\xi+\pi^{T}_{\chi}
\tau^{l}\chi)+\frac{3\gamma\kappa}{8\alpha}
\left(1-\frac{\gamma}{\alpha}\right)
(\pi^{T}_{\xi}\xi-\pi^{T}_{\chi}\chi)(
\pi^{T}_{\xi}\xi-\pi^{T}_{\chi}\chi)
\end{equation}
as it multiplies $q^{-1/2}$. 

\subsection{Equation of state}

From the Hamiltonian we can determine energy and pressure and
formulate the equation of state.  The matter Hamiltonian is directly
related to energy density by
\begin{equation}
\label{rhogeneral}
\rho = \frac{1}{\sqrt{q}}\frac{\delta H_{\rm Dirac}}{\delta N}
\end{equation}
and thus, from (\ref{dirachamiltonianconstraint3}), the energy
density is
\begin{eqnarray}
\label{rho}
\rho &=&
\frac{2E^{a}_{i}}{q}
\left(-\frac{\beta}{2}{\cal D}_{a}\left(\pi^{T}_{\xi}\tau^{i}\xi+
\pi^{T}_{\chi}\tau^{i}\chi\right)+i\left(-\theta_L\pi^{T}_{\xi}\tau^{i}{\cal D}_{a}\xi
    +\theta_R\pi^{T}_{\chi}\tau^{i}{\cal D}_{a}\chi - c.c.\right)\right)
+\frac{B}{q}
\end{eqnarray}
The canonical formula for pressure is
\begin{equation}
\label{pressure}
 P=-{\frac{2}{3N\sqrt{q}}}E^{a}_{i}
 {\frac{\delta{H_{\rm Dirac}}}{\delta{E^{a}_{i}}}}
\end{equation}
as shown by a straightforward adaptation of the calculation done in
\cite{MaxwellEOS} for metric variables. Now using the functional derivative
\begin{equation}
\label{qderivative}
 \frac{\delta \sqrt{q(x)}}{\delta E_{i}^{a}(y)} =
 \frac{1}{2}e^i_a\delta(x-y)\,,
\end{equation}
and thus
\begin{eqnarray}
\label{qderivative1}
 \frac{\delta }{\delta
 E_{j}^{b}(y)}\left(\frac{2E_{i}^{a}(x)}{\sqrt{q(x)}}
 \right) = \frac{1}{\sqrt{q}} (2\delta^{a}_{b}\delta^{j}_{i}-
e^a_ie^j_b)\delta(x-y)\,,
\end{eqnarray}
and inserting (\ref{qderivative1}) in (\ref{pressure}), 
we obtain the pressure
\begin{eqnarray}
\label{pressure2}
P&=& \frac{2E^{a}_{i}}{3q}\left(-\frac{\beta}{2}{\cal
D}_{a}\left(\pi^{T}_{\xi}\tau^{i}\xi+\pi^{T}_{\chi}\tau^{i}\chi\right)+
i\left(-\theta_L\pi^{T}_{\xi}\tau^{i}{\cal D}_{a}\xi
+\theta_R\pi^{T}_{\chi}\tau^{i}{\cal D}_{a}\chi - c.c.\right)\right)
+\frac{B}{q}
\end{eqnarray}
This results in an equation of state
\begin{eqnarray}
\label{eos}
w_{\rm Dirac}= \frac{P}{\rho}=\frac{1}{3}- \frac{2B}{3\rho} \,.
\end{eqnarray}

In relativistic regimes, the kinetic term involving partial
derivatives $\partial_a$ contained in ${\cal D}_a$ is dominant, which
leaves us with an equation of state
\begin{equation}
 w= \frac{P}{\rho}= \frac{1}{3}+\epsilon
\end{equation}
whose leading term agrees with the parameter for a Maxwell field. But
there are clearly correction terms for fermions already in the
classical first order theory.  They do not arise for the Maxwell
field, implying a difference in the coupling to gravity due to
torsion, which is present even in relativistic regimes. The order of
magnitude of the additional term depends on the fermion current
density and is thus not expected to be large unless regimes are very
dense. We will not consider this correction further in this article,
but highlight its role as a consequence of torsion.

In addition, the canonical analysis performed here provides the stage
for a quantization of the theory, resulting in further correction
terms from quantum gravity as they occur even for the Maxwell field
\cite{MaxwellEOS}. Their magnitudes must also be extracted and then
compared for the different fields.

\section{Quantum corrections}
\label{sec:MODIFICATIONS}

We are now in a position to derive the form of quantum gravity
corrections for fermion fields in an explicit scenario of quantum
gravity, completing the program of Sec.~\ref{sec:physics}. We will not
require the quantized dynamics of gravity, which makes our aim
feasible. But we do need to know how the triad is quantized because it
appears in the Dirac Hamiltonian. It appears as a basic variable in
the Poisson relations (\ref{Poisson}), together with the
Ashtekar--Barbero connection $A_a^i$. Loop quantum gravity
\cite{Rov,ALRev,ThomasRev} proceeds by forming a quantum
representation of these basic fields which turns holonomies $h_e(A)=
{\cal P}\exp(\int_e\md t \dot{e}^aA_a^i\tau_i)$ along arbitrary
spatial curves $e$ into creation operators of geometrical excitations
along these curves. Many such creations then excite a discrete state
based on the graph formed by all curves used. On such a state, $E^a_i$
is quantized through the fluxes $F_S(E)= \int_S\md^2y E^a_in_a\tau^i$
through surfaces $S$ with co-normal $n_a$. Resulting operators have
eigenvalues obtained by counting intersections of the surface and the
graph in the state, weighed by multiplicities of overlapping curves.

This structure is important for a background independent
representation which replaces the usual Fock representation of quantum
fields on a curved background space-time. The Fock representation
requires a metric to define basic concepts such as a vacuum state or
creation and annihilation operators, which can thus not be used for
the non-perturbative gravitational field itself. This mathematical
property has important physical implications because extended objects,
namely the integrated holonomies and fluxes, are used as basic
quantities. In this way, there is non-locality and other implications
which affect quantum gravity corrections.

In loop quantum gravity, there are three main effects which imply
correction terms in effective matter equations. Any Hamiltonian
contains inverse powers of triad components such as $1/\sqrt{q}$ in
(\ref{dirachamiltonianconstraint3}), which for fermions is a
consequence of the fact that they are quantized through half-densities
\cite{FermionHiggs}. The loop quantization, however, leads to triad
operators which have discrete spectra containing zero, thus lacking
inverse operators. A proper quantization, along the lines of
\cite{QSDV}, does give a well-defined operator with the correct
semiclassical limit. But there are deviations from the classical
behavior on small length scales, which are the first source of
correction terms. As in the case of the Maxwell field
\cite{MaxwellEOS}, this is the main effect we include here.

In addition, there are qualitatively different correction
terms. First, loop quantum gravity is spatially discrete, with states
supported on spatial graphs. Quantizations of Hamiltonians thus lead
to a discrete representation of any spatial derivative term as they
also occur for fermions. The classical expression arises in a
continuum limit, but for any given state the discrete representation
implies corrections to the classical derivatives as the leading terms
in an expansion. Secondly, the connection is quantized through
holonomies rather than its single components. Thus, the quantum
Hamiltonians are formulated in terms of exponentials of line integrals
of the connection which also give the leading classical term plus
corrections in an expansion. Finally, whenever a Hamiltonian is not
quadratic, there are genuine quantum effects as they occur in typical
low energy effective actions. They can be computed in a Hamiltonian
formulation as well \cite{EffAc,Karpacz}, contributing yet another source of
corrections.

One certainly needs to know the relative magnitude of all corrections
in order to see which ones have to be taken into account. For all of
them, the magnitude depends on details of the quantum state describing
the regime.  Here, properties of states have to be taken into account,
and dimensional arguments are no longer sufficient.  For instance,
discretization and curvature corrections depend on the patch size
occurring in the discrete state underlying quantum gravity. This patch
size is typically small compared to scales on which the matter field
changes, even in relativistic regimes assumed here. Thus, such
corrections can be ignored in a first approximation. What remains are
corrections from inverse powers. While other corrections shrink in the
continuum limit where the patch size becomes small, inverse
corrections actually grow when the patch size approaches the Planck
length. The regimes where the two classes of corrections are dominant
are thus neatly separated, and we can safely focus on inverse triad
corrections only. The relevant formulas are collected in Appendix
\ref{appen:lqg}; see also \cite{QuantCorrPert}. A detailed and
complete derivation is not yet available since precise properties of a
quantum gravity state would be required. Still, many general
qualitative insights can be gained in this way.

In the Dirac Hamiltonian (\ref{dirachamiltonianconstraint3}) the
factor to be quantized containing inverse powers of the densitized
triad is
\[
 \frac{2E^{a}_{i}}{\ell_0\sqrt{q}}=
 \epsilon^{abc}\epsilon_{ijk}\frac{e_b^je_c^k}{\ell_0\sqrt{q}} \approx
 \epsilon^{abc}\epsilon_{ijk}\frac{\ell_0^2 e_b^je_c^k}{V_v}
\]
in terms of the volume $V_v\approx\ell_0^3\sqrt{q(v)}$ of one discrete
patch at a point $v$.  We can already notice the close resemblance to
the Maxwell Hamiltonian, where the corresponding expression is
$q_{ab}/\ell_0\sqrt{q}= e_a^ie_b^i/\ell_0\sqrt{q}$ which differs only
by the additional $\epsilon$-tensors. This close relation will, in the
end, lead to quite similar quantum corrections for photons and
fermions.

We proceed using (\ref{AV}) for $r=1/2$, and write
\begin{equation}
\label{AV1}
 \frac{2E^{a}_{i}}{\ell_0\sqrt{q}}=
 \left(\frac{\ell_0}{2\pi G \gamma}\right)^2
 \epsilon^{abc}\epsilon_{ijk}\{A_b^j,V_v^\frac{1}{2}\}
\{A_c^k,V_v^\frac{1}{2}\},
\end{equation}
which can then be quantized by turning Poisson brackets into
commutators of operators. This results in 
\begin{equation}
\widehat{\frac{2E^{a}_{i}}{\ell_0\sqrt{q}}}=
\epsilon^{KIJ}\epsilon_{ijk}\widehat{(\ell_0V_v^{-1/2}e_I^j)} 
\widehat{(\ell_0V_v^{-1/2}e_J^k)}
= \epsilon^{KIJ}\epsilon_{ijk}\hat{C}_{v,I}^{(1/2)}
\hat{C}_{v,J}^{(1/2)} \delta^j_{(I)}\delta^k_{(J)}
\end{equation}
with $\hat{C}_{v,I}^{1/2}$ defined in (\ref{BC}).  As explicit
calculations in the appendix show, eigenvalues (\ref{Cv}) of the
operator $\hat{C}_{v,I}^{(1/2)}$, or its expectation values in
semiclassical states, do not follow precisely the behavior expected
for the classical function it quantizes. Deviations become larger for
small values of the triad components, which can be captured in a
correction function $\alpha[p^I(v)]$ multiplying the classical
expression. In particular, such a correction function will appear in a
Hamiltonian operator, thus also correcting expressions for energy
density and pressure or the equation of state. The functional form
follows from Eqs.~(\ref{Cv}) and (\ref{alphav}).

Thus the general expression one can expect for a phenomenological
Dirac Hamiltonian including corrections from inverse powers of the
triad is
\begin{eqnarray}
\label{dirachamiltonianconstraint5}
 H_{\rm phen}
&=& \int_{\Sigma_{t}} \md^{3}xN\left(
   \frac{E^{a}_{i}}{\sqrt{q}}\alpha(E^{b}_{j})\left(-\beta
{\cal
   D}_{a}\left(\pi^{T}_{\xi}\tau^{i}\xi+\pi^{T}_{\chi}
\tau^{i}\chi\right)-2i
\left(\theta_{L}\pi^{T}_{\xi}\tau^{i}{\cal D}_{a}\xi 
 -\theta_{R}\pi^{T}_{\chi}\tau^{i}{\cal D}_{a}\chi -
 c.c.\right)\right)\right.
+\left.\frac{\theta(E^{b}_{j})}{\sqrt{q}} B\right)
\end{eqnarray}
with two possibly different correction functions $\alpha$ and
$\theta$.  This also affects the energy density and pressure terms,
derived by the general expressions (\ref{rhogeneral}) and
(\ref{pressure}).  We are mainly interested in the correction to the
one-third in the equation of state (\ref{eos}), so we focus on the
first term in (\ref{dirachamiltonianconstraint5}) in what
follows. Moreover, for a {\em nearly} \footnote{We are not assuming
strict isotropy to compute quantum corrections of inhomogeneous
fermion fields. Nevertheless, in leading order corrections one can use
the background geometry.}  isotropic background geometry $\alpha$ only
depends on the determinant $q$ of the spatial metric and thus
$q^{ab}\delta\alpha/\delta q^{ab}= - 3q\md\alpha/\md
q=-\frac{1}{2}a\md\alpha/\md a$ with the scale factor $a$ related to
$q$ by $q=\det(q_{ab})=a^6$. In this case the quantum gravitational
expectation for $\alpha(q)$, as per Eqs.~(\ref{Cv}) and
(\ref{alphav}), simplifies.  To use these expressions, we have to
relate the scale factor to quantum gravitational excitation levels as
they occur in calculations of loop quantum gravity.  In the notation
of the appendix, an elementary discrete patch in a nearly isotropic
space-time has, on the one hand, an area of $\ell_0^2a^2$ if $\ell_0$
is the coordinate diameter of the patch. This can be expressed as
$\ell_0^2a^2= (V_{\cal V}/{\cal N}_{\cal V})^{2/3}$ where ${\cal
N}_{\cal V}$ is the number of patches in a box ${\cal V}$ of volume
$V_{\cal V}$. On the other hand, using (\ref{FluxVert}) the quantum
gravity state assigns a value of $4\pi\gamma\ell_{\rm P}^2\mu_v$ to
this patch via the flux operator, where $\mu_v$ is the quantum number
of the geometrical excitation of this patch. Thus, we obtain
\[
 \mu_v = \frac{V_{\cal V}^{2/3}}{4\pi\gamma\ell_{\rm P}^2 {\cal
 N}_{\cal V}^{2/3}}=:\frac{a^2}{a_{\rm disc}^2}
\]
where
\begin{equation}
 a_{\rm disc} = 2\sqrt{\pi\gamma} \ell_{\rm P} \left(\frac{{\cal
 N}_{\cal V}}{V_0}\right)^{1/3}
\end{equation}
with the coordinate volume $V_0$ of the box ${\cal V}$. The numerical
value of $a_{\rm disc}$ depends on coordinates via $V_0$, or on the
normalization of the scale factor. (It does not depend on the choice
of the box ${\cal V}$ because a change would multiply ${\cal N}_{\cal
V}$ and $V_0$ by the same factor.) But it is important to note that
$a_{\rm disc}$ is not just determined by the Planck length $\ell_{\rm
P}$, which appears for dimensional reasons, but also depends on the
large number ${\cal N}_{\cal V}$ of discrete patches per volume as
given by the quantum gravity state. This is exactly a parameter as
expected in the discussion of Sec.~\ref{sec:physics}.  Replacing
$\mu_v$ in the equations of the appendix, we obtain
\begin{equation}\label{alphafull}
 \alpha(a)=8\sqrt{2}(a/a_{\rm disc})^2 \left((2(a/a_{\rm
     disc})^2+1)^{1/4}- |2(a/a_{\rm disc})^2-1|^{1/4}\right)^2
\end{equation}
where $a_{\rm disc}$ appears, influencing the size of quantum gravity
corrections.

\begin{figure}[t]
\begin{center}
\includegraphics[width=12cm]{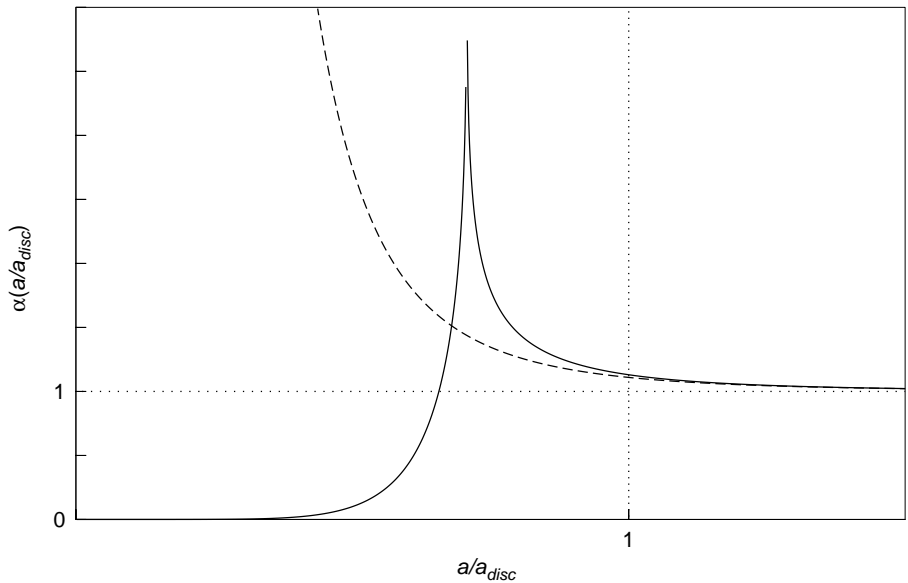}
\end{center}
\caption{The correction function (\ref{alphafull}) as a function of
  the scale factor (solid line). The asymptotic form (\ref{alphasim})
  for large $a$ is shown by the dashed line. (The sharp cusp, a
  consequence of the absolute value appearing in (\ref{alphafull}), is
  present only for eigenvalues as plotted here, but would disappear
  for expectation values of the inverse volume operator in coherent
  states. This cusp will play no role in the analysis of this paper.)}
\label{Fig:alpha}
\end{figure}

The function is plotted in Fig.~\ref{Fig:alpha}.  One can easily see
that $\alpha(a)$ approaches the classical value $\alpha=1$ for $a\gg
a_{\rm disc}/\sqrt{2}$, while it differs from one for small $a$. For
$a>a_{\rm disc}/\sqrt{2}$, the corrections are perturbative in
$a^{-1}$,
\begin{equation} \label{alphasim}
 \alpha(a)\sim 1+\frac{7}{64}\left(\frac{a_{\rm disc}}{a}\right)^4+\cdots\,.
\end{equation}
This is the first correction in an asymptotic expansion for
eigenvalues. If semiclassical states rather than volume eigenstates
are used, powers of $a^{-1}$ in the leading corrections can be
smaller.  Moreover, via ${\cal N}_{\cal V}$ the discreteness scale
$a_{\rm disc}$ is expected to be not precisely constant but a function
of $a$ itself because the underlying spatial discreteness of quantum
gravity can be refined dynamically during cosmological evolution
\cite{InhomLattice,CosConst}. (Indeed, dynamical refinement is also
required for several other phenomenological reasons
\cite{APSII,SchwarzN,RefinementInflation,RefinementMatter,Tensor}.)
In our following analysis we will thus assume a functional form
\begin{equation}
\label{alpha(a)}
\alpha(a) = 1 + c(a/a_0)^{-n}
\end{equation}
where we traded the fundamental normalization by $a_{\rm disc}$ for
normalization with respect to the present-day value of the scale
factor $a_0$. From the derivation, $n$ is likely to be a small, even
integer and $c$ is known to be positive.  The constant $c$ depends on
$a_{\rm disc}$ and inherits the ${\cal N}_{\cal V}$-factor. It can
thus be larger than of order one. We will treat this parameter as
phenomenological and in the end formulate bounds on $c$ as bounds for
${\cal N}_{\cal V}$.

Energy density and the pressure then are, ignoring the classical
interaction term $B$,
\begin{eqnarray}
\label{rhoeffective}
 \rho_{\rm eff}
&=&  \frac{2E^{a}_{i}}{a^6} \alpha(a)
\left(-\beta{\cal D}_{a}\left(\pi^{T}_{\xi}\tau^{i}\xi+
\pi^{T}_{\chi}\tau^{i}\chi\right)+i\left(-\theta_L\pi^{T}_{\xi}\tau^{i}
{\cal D}_{a}\xi
    -\theta_R\pi^{T}_{\chi}\tau^{i}{\cal D}_{a}\chi - c.c.\right)\right)
\end{eqnarray}
and
\begin{eqnarray}
\label{pressureeffective}
 3P_{\rm eff}
&=& \frac{2E^{a}_{i}}{a^6}\alpha(a)\left(1-\frac{{\rm d \;
      log} 
\; \alpha}{{\rm d \;log} \;a}\right)
\left(-\beta{\cal D}_{a}
\left(\pi^{T}_{\xi}\tau^{i}\xi+\pi^{T}_{\chi}\tau^{i}\chi\right)+
i\left(-\theta_L\pi^{T}_{\xi}\tau^{i}{\cal D}_{a}\xi
 -\theta_R\pi^{T}_{\chi}\tau^{i}{\cal D}_{a}\chi - c.c.\right)\right)\,.
\end{eqnarray}
From this, the equation of state $w$ can easily be computed:
\begin{eqnarray}
\label{modifiedw}
w_{\rm eff} = \frac{1}{3}\left(1 - \frac{{\rm d \;log}\; \alpha}{{\rm d
\;log}\; a}\right)\,.
\end{eqnarray}
This quantum gravity correction is independent of the specific matter
dynamics as in the classical relativistic case. It results in an
equation of state which is linear in $\rho$, but depends on the
geometrical scales (and the Planck length) through $\alpha$. This is
the same general formula derived in \cite{MaxwellEOS} for radiation.
Thus, on an isotropic background radiation and relativistic fermions
are not distinguished by the form of quantum corrections they receive.

With the equation of state and the continuity equation
\begin{equation}
\label{continuity}
\dot{\rho} + 3\frac{\dot{a}}{a}\left(\rho+P\right)= 0\,,
\end{equation}
where $a$ is the scale factor and the dot indicates a proper time
derivative, we can determine energy density as a function of the
universe size. We first obtain
\begin{equation}
\label{usefulrho}
 \frac{\md\log \rho (a)}{\md\log
 a} = - 3 \left(1 + w(a)\right)\,,
\end{equation}
explicitly showing the dependence of the equation of state on the
scale factor. The solution is
\begin{equation}
\label{solutionrho}
\rho (a)= \rho_{0}
\exp\left[-3\int\left(1+w(a)\right)\md\log a\right],
\end{equation}
where $\rho_{0}$ is the integration constant. Inserting 
an equation of state of the form (\ref{modifiedw})
we obtain
\begin{equation}
\label{correctedrho}
\rho(a)= \rho_{0} \alpha(a) a^{-4}.
\end{equation}
For $\alpha$ = 1, we retrieve the classical result $\rho(a)\propto
a^{-4}$, but for $\alpha\not=1$ loop quantum gravity corrections
induced by the discreteness of flux operators are reflected in the
evolution of energy density in a Friedmann--Robertson--Walker
universe.

\section{Effect on Big Bang Nucleosynthesis}
\label{sec:BBN}

The production of elements in the early universe is highly sensitive
to the expansion rate, given by
\begin{equation}
\frac{\dot{a}}{a} = \left(\frac{8}{3}\pi G \rho \right)^{1/2},
\end{equation}
where $\rho$ is the total density, thus including radiation and
fermions.  As we have seen here for fermions and in \cite{MaxwellEOS}
for radiation, the effect of loop quantum gravity corrections is to
multiply the effective $\rho(a)$ by a factor $\alpha(a)$.  Most
importantly, we find that $\alpha(a)$ is the {\it same} for both
bosons and fermions (up to possible quantization ambiguities), so a
separate treatment of the two types of particles in the early universe
(as in Ref. \cite{FermionBoson}) is unnecessary here.

In the standard treatment of the thermal history of the universe,
the density of relativistic particles (bosons or fermions) is given
by
\begin{equation}
\label{rhostandard}
\rho = \frac{\pi^2}{30}g_* T^4,
\end{equation}
where $g_*$ is the number of spin degrees of freedom for bosons,
and $7/8$ times the number of spin degrees of freedom for fermions,
and $T$ is the temperature, which scales as
\begin{equation}
\label{Tstandard}
T\propto a^{-1}.
\end{equation}
The equation of state parameter is
\begin{equation}
\label{pstandard}
w = 1/3.
\end{equation}
Clearly, equations (\ref{rhostandard})--(\ref{pstandard}) are inconsistent
with equations (\ref{modifiedw}) and (\ref{correctedrho}).  There is
some ambiguity in determining the correct way to modify the expressions
for $\rho(T)$ and $w$.  We have chosen to assume that the modifications
are contained in the gravitational sector, so that the density
is given by
\begin{equation}
\rho = \alpha(a)\frac{\pi^2}{30}g_* T^4,
\end{equation}
with the temperature scaling as in equation (\ref{Tstandard}), and the
equation of state $w$ is given by equation (\ref{modifiedw}).  This
guarantees that the standard continuity equation (\ref{continuity})
continues to hold.  Note that this is not the only way to incorporate
equation (\ref{correctedrho}) into the calculation, but it seems to us
the most reasonable way.  This issue requires a consideration of
thermodynamics on a quantum space-time, which is a fascinating but not
well-studied area. Instead of entering details here, we note that we
interpret the $\alpha$-correction as a consequence of a quantum
gravity sink to energy and entropy. Thus, quantum gravity implies a
non-equilibrium situation which would otherwise imply that $\rho$ must
be proportional to $T^4$ without any additional dependence on
$a\propto T^{-1}$.

With these assumptions, we can simply treat $\alpha(a)$ as an
effective multiplicative change in the overall value of $G$.  Note
that this simplification is only possible because we explicitly
derived by our canonical analysis that, unexpectedly, quantum
corrections of radiation and fermions appear in similar forms. This
makes possible a comprehensive derivation of implications for BBN,
bearing on earlier work.  In fact, a great deal of work has been done
on the use of BBN to constrain changes in $G$ (see, e.g.
Refs.~\cite{Barrow,Yang,Accetta,Copi,Clifton}).  The calculation is
straightforward, if one has a functional form for the time-variation
in $G$.  For the loop quantum gravity corrections considered here, the
most reasonable functional form is (\ref{alpha(a)}). Note that this
expression is by construction valid only in the limit where $\alpha(a)
- 1 << 1$.  In terms of the effective gravitational constant, $G$, one
can then write
\begin{equation}
\label{G(a)}
G(a) = G_0[1 + c(a/a_0)^{-n}],
\end{equation}
where $G_0$ is the present-day value of the gravitational
constant.

In order to constrain the values of $c$ and $n$, we calculate
the predicted element abundances with the indicated change in
$G$ and compare with observational constraints.  Big Bang
nucleosynthesis proceeds first through the weak interactions that
interconvert protons and neutrons:
\begin{eqnarray}
n+\nu _{e} &\leftrightarrow &p+e^{-},  \nonumber \\
n+e^{+} &\leftrightarrow &p+\bar{\nu}_{e},  \nonumber \\
n &\leftrightarrow &p+e^{-}+\bar{\nu}_{e}.
\end{eqnarray}
When $T\gtrsim 1$ MeV, the weak-interaction rates are faster than the
expansion rate, $\dot{a}/a$, and the neutron-to-proton ratio ($n/p$)
tracks its equilibrium value $\exp [-\Delta m/T]$, where $\Delta m$ is
the neutron-proton mass difference. As the universe expands and cools,
the expansion rate becomes too fast for weak interactions to maintain
weak equilibrium and $n/p$ freezes out. Nearly all the neutrons which
survive this freeze-out are converted into $^4$He as soon as deuterium
becomes stable against photodisintegration, but trace amounts of other
elements are produced, particularly deuterium and $^7$Li (see, e.g.,
Ref.~\cite{OSW} for a review).

In the standard model, the predicted abundances of all of these
light elements are a function of the baryon-photon ratio, $\eta$, but
any change in $G$ alters these predictions.
Prior to the era of precision CMB observations (i.e., before WMAP),
Big Bang nucleosynthesis provided the most stringent constraints on $\eta$,
and modifications to the standard model could be ruled out only
if no value of $\eta$ gave predictions for the light element abundances
consistent with the observations.  However, the CMB observations
now provide an independent estimate for $\eta$, which can be used
as an input parameter for Big-Bang nucleosynthesis calculations.

Copi et al. \cite{Copi} have recently argued that the most reliable
constraints on changes in $G$ can be derived by using the WMAP values
for $\eta$ in conjuction with deuterium observations.  The reason is
that deuterium can be observed in (presumably unprocessed)
high-redshift quasi stellar object (QSO) absorption line systems (see
Ref.~\cite{Kirkman} and references therein), while the estimated
primordial $^4$He abundance, derived from observations of low
metallicity HII regions, is more uncertain (see, for example, the
discussion in Ref.~\cite{Olive}).  While we agree with the argument of
Copi et al. in principle, for the particular model under consideration
here it makes more sense to use limits on $^4$He than on deuterium, in
conjunction with the WMAP value for $\eta$.  The reason is that the
$^4$He abundance is most sensitive to changes in the expansion rate at
$T \sim 1$ MeV, when the freeze-out of the weak interactions
determines the fraction of neutrons that will eventually be
incorporated into $^4$He.  Deuterium, in contrast, is produced in Big
Bang nucleosynthesis only because the expansion of the universe
prevents all of the deuterium from being fused into heavier elements.
Thus, the deuterium abundance is most sensitive to the expansion rate
at the epoch when this fusion process operates ($T \sim 0.1$ MeV).
The importance of this distinction with regard to modifications of the
standard model was first noted in Ref.~\cite{Kolb}, and a very nice
quantitative analysis was given recently in Ref.~\cite{Bambi}.  Note
that our estimate for the behavior of $G(a)/G_0 - 1 $, equation
(\ref{G(a)}), is a steeply decreasing function of $a$.  Thus, the
change in the primordial $^4$He abundance will always be much larger
than the change in the deuterium abundance.  Therefore, we can obtain
better constraints on this model by using extremely conservative
limits on $^4$He, rather than by using the more reliable limits on the
deuterium abundance.  For the same reason, we can ignore any effect on
the CMB, since the latter is generated at a much larger value of $a$,
and any change will be minuscule.  Hence, we can confidently use the
WMAP value for $\eta$.

\begin{figure}[t]
\centerline{\epsfxsize=3.3truein\epsffile{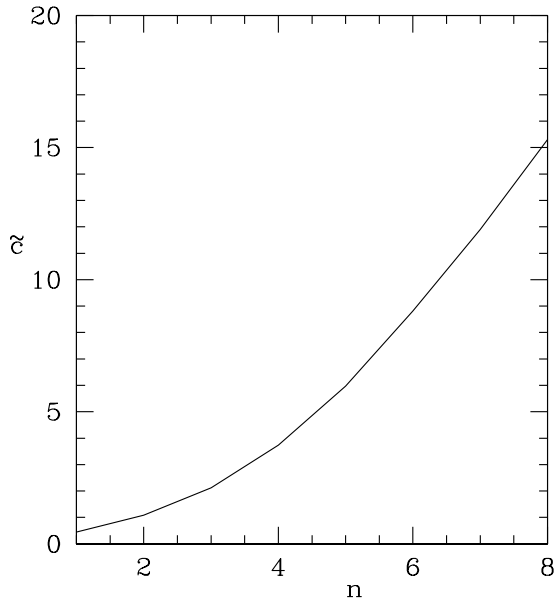}}
\caption{Solid curve gives upper bound on $\widetilde c$ as a function
of $n$, for the assumed form for $\alpha$: $\alpha = 1 + \widetilde
c/a_{10}^n$, where $a_{10}$ is the value of the scale factor in units
for which $a_{10} = 10^{10}$ at present.}
\label{Fig:Bounds}
\end{figure}

WMAP gives \cite{Spergel}
\begin{equation}
\label{eta}
\eta = 6.116^{+0.197}_{-0.249} \times 10^{-10}.
\end{equation}
Because the estimated errors on $\eta$ are so small, we simply use the central
value for $\eta$; the bounds we derive on $c$ in equation
(\ref{alpha(a)}) change only slightly when $\eta$ is varied within
the range given by equation (\ref{eta}).  Since $c$ in equations
(\ref{alpha(a)}) and (\ref{G(a)}) is thought to be positive, the effect
of LQG corrections is to increase the primordial
expansion rate, which increases the predicted $^4$He abundance.
We therefore require an observational upper bound on the primordial
$^4$He abundance.  As noted earlier, this is a matter of some
controversy.  We therefore adopt the very conservative upper
bound recommended by Olive and Skillman \cite{Olive}:
\begin{equation}
Y_P \le 0.258,
\end{equation}
where $Y_P$ is the primordial mass fraction of $^4$He.  For a fixed
value of $n$ in equation (\ref{G(a)}), we determine the largest value
of $c$ that yields a primordial $^4$He abundance consistent with this
upper limit on $Y_P$.  Since we are essentially bounding the
change in $G$ at $a/a_0 \sim 10^{-10}$, it is convenient to rewrite
equation (\ref{alpha(a)}) as
\begin{equation}
\label{alpha(a)2}
\alpha = 1 + \widetilde c/a_{10}^n
\end{equation}
where $a_{10} \equiv 10^{10}(a/a_0)$.
This upper bound on $\widetilde c$ as a function of $n$
is given in Fig.~\ref{Fig:Bounds}.  For the special case $n=4$, we
can use these results to place a bound on $a_{\rm disc}$ in equation
(\ref{alphasim}).  We obtain
\begin{equation}
\frac{a_{\rm disc}}{a_0} < 2.4 \times 10^{-10}\,.
\end{equation}
This is not a strong bound for the parameters of quantum gravity, but
clearly demonstrates that quantum corrections are consistent with
successful big bang nucleosynthesis.

In terms of more tangible quantum gravity parameters, we have
\begin{equation}
 {\cal N}_{\cal V}^{1/3} < \frac{1.2\cdot 10^{-10}}{\sqrt{\pi\gamma}}
 \frac{a_0 V_0^{1/3}}{\ell_{\rm P}}
\end{equation}
for the number of patches at the time of big bang nucleosynthesis. In
terms of the volume $V_{\cal V}=(10^{-10}a_0)^3 V_0$ at this time, we
have ${\cal N}_{\cal V}<3 V_{\cal V}/\ell_{\rm P}^3$ with the value
$\gamma\approx 0.24$ of the Barbero--Immirzi parameter as derived from
black hole entropy calculations.  More meaningfully, if we view
$2\sqrt{\pi\gamma}\ell_{\rm P}$ as the basic length scale as it
appears in the spectrum (\ref{FluxVert}) of loop quantum gravity, the
bound becomes more interesting: This gives ${\cal N}_{\cal
V}^{1/3}<2.4 V_{\cal V}^{1/3}/(2\sqrt{\pi\gamma}\ell_{\rm P})$.  This
upper limit is already quite close to what one expects for elementary
patch sizes in loop quantum gravity, which would provide ${\cal
N}_{\cal V}^{1/3}< V_{\cal V}^{1/3}/(2\sqrt{\pi\gamma}\ell_{\rm P})$
as a fundamental upper limit.  Given that these values are close to
each other, we see a clear potential of improvements by more precise
observational inputs. Moreover, other correction terms from quantum
gravity could be used to obtain a lower bound for ${\cal N}_{\cal V}$
such that the allowed window would be reduced to a smaller size.

\section{Conclusions}

Big bang nucleosynthesis is a highly relativistic regime which, to a
good approximation, implies identical equations of state for fermions
and photons. There are, however, corrections to the simple equation of
state $w=\frac{1}{3}$ for fermions even classically. One observation
made here is that the interaction term derived in
\cite{FermionImmirzi} leads to such a correction and might be more
constrained by nucleosynthesis than through standard particle
experiments \cite{FermionTorsion}. We have not analyzed this further
here because more details of the behavior of the fermion current would
be required.

A second source of corrections arises from quantum
gravity. Remarkably, while quantum gravity effects on an isotropic
background do correct the equations of state, they do so equally for
photons and relativistic fermions. Initially, this is not expected for
both types of fields due to their very different actions. Thus,
quantum gravity effects do not spoil the detailed balance required for
the scenario to work and bounds from big bang nucleosynthesis obtained
so far are not strong.  But there are interesting limits for the
primary parameter, the patch size of a quantum gravity state. It is
dimensionally expected to be proportional to the Planck length
$\ell_{\rm P}$ but could be larger. In fact, current bounds derived
here already rule out a patch size of exactly the elementary allowed
value in loop quantum gravity. With more precise estimates, these
bounds can be improved further.

We have made use of quantum gravity corrections in a form which does
not distinguish fermions from radiation. Although the most natural
implementation, quite unexpectedly, provides equal corrections as
shown here, there are several possibilities for differences which
suggest several further investigations.  Small deviations in the
equations of state and thus energy densities of fermions and radiation
are possible. First, there are always quantization ambiguities, and so
far we tacitly assumed that the same basic quantization choice is made
for the Maxwell and Dirac Hamiltonians.  Such ambiguity parameters can
be explicitly included in specific formulas for correction functions;
see e.g.\ \cite{Ambig,ICGC,QuantCorrPert}. Independent consistency
conditions for the quantization may at some point require one to use
different quantizations for both types of fields, resulting in
different quantum corrections and different energy densities. Such
conditions can be derived from an analysis of anomaly-freedom of the
Maxwell field and fermions coupled to gravity, which is currently in
progress. As shown here, if this is the case it will become testable
in scenarios sensitive to the behavior of energy density such as big
bang nucleosynthesis.  Moreover, assuming the same quantization
parameters leads to identical quantum corrections for photons and
fermions only on isotropic backgrounds. Small-scale anisotropies have
different effects on both types of fields and can thus also be probed
through their implications on the equation of state.

For this, it will be important to estimate more precisely the typical
size of corrections, which is not easy since it requires details of
the quantum state of geometry. The crucial ingredient is again the
patch size of underlying lattice states. On the other hand, taking a
phenomenological point of view allows one to estimate ranges for patch
sizes which would leave one in agreement with big bang nucleosynthesis
constraints.  Interestingly, corrections studied here provide upper
bounds to the patch size, and other corrections from quantum gravity
are expected to result in lower bounds. A finite window thus results,
which can be shrunk with future improvements in observations.

\begin{acknowledgments}

  M.B. was supported in part by NSF grants PHY05-54771 and PHY06-53127.
  R.J.S. was supported in part by the Department of Energy (DE-FG05-85ER40226).
  Some of the work was done while M.B. visited the
  Erwin-Schr\"odinger-Institute, Vienna, during the program ``Poisson Sigma
  Models, Lie Algebroids, Deformations, and Higher Analogues,'' whose
  support is gratefully acknowledged.
\end{acknowledgments}

\begin{appendix}

\section{Elements of loop quantum gravity}
\label{appen:lqg}

We collect the basic formulas required to compute quantum corrections
in loop quantum cosmology, referring to
\cite{QuantCorrPert,MaxwellEOS} for further details.  We can restrict
our attention to perturbative regimes around a spatially flat
isotropic solution where one can choose the canonical variables to be
given by functions $(\tilde{p}^I(x),\tilde{k}_J(x))$ which determine a
densitized triad by $E^a_i=\tilde{p}^{(i)}(x)\delta^a_i$ and extrinsic
curvature by $K_a^i=\tilde{k}_{(i)}(x)\delta_a^i$. This
diagonalization implies strong simplifications in the explicit
evaluation of formulas \cite{CosmoI,SphSymm}.

Any state is associated with a spatial lattice with U(1) elements
$\eta_{v,I}$ attached to its links $e_{v,I}$ starting at a vertex $v$
and pointing in a direction along a translation generator $X_I^a$ of
the homogeneous background.  In our context of perturbations around
Friedmann--Robertson--Walker space-times, this can be seen as a
definition of the quantum gravity state on which the quantized $E^a_i$
and $K_a^i$ act.  The U(1) elements $\eta_{v,I}$ appear as matrix
elements in SU(2) holonomies $h_{v,I}={\rm Re}\eta_{v,I}+2\tau_I {\rm
Im}\eta_{v,I}$ and represent the connection, while fluxes $F_{v,I}$
are used as independent variables for the momenta.  An orthonormal
basis of the quantum Hilbert space is given by functionals
$|\ldots,\mu_{v,I},\ldots\rangle=\prod_{v,I} \eta_{v,I}^{\mu_{v,I}}$,
for all possible choices of $\mu_{v,I}\in{\mathbb Z}$. Together with
basic operators, which are represented as holonomies
\begin{equation}
 \hat{\eta}_{v,I}|\ldots,\mu_{v',J},\ldots\rangle =
|\ldots,\mu_{v,I}+1,\ldots\rangle
\end{equation}
for each pair $(v,I)$, and fluxes
\begin{equation} \label{FluxVert}
 \hat{{\cal F}}_{v,I} |\ldots,\mu_{v',J},\ldots\rangle=
 2\pi\gamma\lP^2(\mu_{v,I}+\mu_{v,-I})
|\ldots,\mu_{v',J},\ldots\rangle\,,
\end{equation}
this defines the basic quantum representation.  Here, $\lP=\sqrt{\hbar
G}$ is the Planck length and a subscript $-I$ means that the edge
preceding the vertex $v$ in the chosen orientation is taken.

For quantum corrections to classical equations, one important step in
addition to computing expectation values of operators is the
introduction of a continuum limit. This relates holonomies
\begin{equation}
 \eta_{v,I} = \exp(i\smallint_{e_{v,I}}\md t
 \gamma\tilde{k}_I/2)\approx \exp(i\ell_0 \gamma\tilde{k}_I(v+I/2)/2)
\end{equation}
to continuum fields $\tilde{k}_I$ through mid-point evaluation
(denoted by $v+I/2$), and similarly for fluxes
\begin{equation} \label{ScalarFlux}
 F_{v,I}=
\int_{S_{v,I}} \tilde{p}^I(y)\md^2y\approx
\ell_0^2\tilde{p}^I(v+I/2)\,.
\end{equation}
Here, $\ell_0$ is the coordinate length of lattice links, which enters
the continuum approximation since we are integrating classical fields
in holonomies and fluxes.
From flux operators one defines the volume
operator $\hat{V}=\sum_{v} \prod_{I=1}^3 \sqrt{|\hat{\cal
F}_{v,I}|}$, using the classical expression $V=\int\md^3x
\sqrt{|\tilde{p}^1\tilde{p}^2\tilde{p}^3|}\approx \sum_v\ell_0^3
\sqrt{|\tilde{p}^1\tilde{p}^2\tilde{p}^3|}= \sum_v\sqrt{|p^1p^2p^3|}$.
With (\ref{FluxVert}), its eigenvalues are
\begin{equation}\label{V_action}
 V(\{\mu_{v,I}\})= \left(2\pi\gamma\lP^2\right)^{3/2} \sum_{v}
 \prod_{I=1}^3\sqrt{ |\mu_{v,I}+\mu_{v,-I}|}\,.
\end{equation}

The volume operator is central for inverse triad corrections because
inverse densitized triads, or a co-triad, can be quantized using
relations such as 
\begin{equation}
\label{AV}
 \{A_a^i,V_v^r\}=4\pi\gamma G\:rV_v^{r-1}e_a^i
\end{equation}
for $0<r<2$. Even though an inverse power of volume, together with a
co-trad, occurs on the right hand side, the left hand side can be
quantized directly in terms of the volume operator, using holonomies
for connection components, and turning the Poisson bracket into a
commutator \cite{QSDI,QSDV}. Such a quantization leads to
\begin{equation} \label{BC}
\widehat{V_v^{r-1}{e}_I^i}= \frac{-2}{8\pi i r\gamma \lP^2 \ell_0}
\sum_{\sigma \in \{\pm 1\}}\sigma\tr(\tau^ih_{v,\sigma I}[h_{v,\sigma
I}^{-1},\hat{V}{}_v^r]) =\frac{1}{2\ell_0} (\hat{B}_{v,I}^{(r)} -
\hat{B}_{v,-I}^{(r)}) \delta^i_{(I)} =:\frac{1}{\ell_0}
\hat{C}_{v,I}^{(r)} \delta^i_{(I)}
\end{equation}
where, for symmetry, we use both edges touching the vertex $v$ along
direction $X_I^a$ and $\hat B_{v,I}^{(r)}$ is, after taking the trace
in (\ref{BC}),
\begin{equation}\label{B_def}
\hat B_{v,I}^{(r)} := \frac{1}{4 \pi i\gamma G \hbar r}\left(s_{v,I}
\hat V_v^r c_{v,I} - c_{v,I}\hat V_v^r s_{v,I}\right)
\end{equation}
with
\[
 c_{v,I}=\frac{1}{2}(\eta_{v,I}+\eta_{v,I}^*) \quad \mbox{ and }\quad
s_{v,I}=\frac{1}{2i}(\eta_{v,I}-\eta_{v,I}^*)\,.
\]

As in \cite{QuantCorrPert} effects of the quantization of triad
(metric) coefficients are included by inserting correction functions
in the classical Hamiltonian which follow, e.g., from the eigenvalues
\cite{QuantCorrPert}
\begin{equation} \label{Cv}
 C_{v,I}^{(1/2)}(\{\mu_{v',I'}\})= 2(2\pi\gamma\ell_{\rm P}^2)^{-1/4}
|\mu_{v,J}+ \mu_{v,-J}|^{1/4} |\mu_{v,K}+ \mu_{v,-K}|^{1/4}
\left(|\mu_{v,K}+ \mu_{v,-K}+1|^{1/4} - |\mu_{v,K}+
\mu_{v,-K}-1|^{1/4} \right)
\end{equation}
(where indices $J$ and $K$ are defined such that
$\epsilon_{IJK}\not=0$) of operators $\hat{C}^{(1/2)}_{v,I}$. 

Classically, we expect $q_{IJ}/\sqrt{q}= \sqrt{|p^1p^2p^3|}/p^Ip^J$
for this quantity, with a densitized triad $E^a_i=p^{(i)}\delta^a_i$
and using the relation (\ref{FluxVert}) between labels and flux
components.  Although for large $\mu_{v,I}$ the eigenvalues indeed
approach the function
\[
C_{v,I}^{(1/2)}(\{\mu_{v',I'}\}) C_{v,J}^{(1/2)}(\{\mu_{v',I'}\})
\sim (2\pi\gamma\ell_{\rm
P}^2)^{-1/2} \frac{\prod_{K=1}^3\sqrt{|\mu_{v,K}+\mu_{v,-K}|}}{
|\mu_{v,I}+\mu_{v,-I}||\mu_{v,J}+\mu_{v,-J}|}
\]
they from the classical expectation
differ for values of $\mu_{v,I}$ closer to one. This deviation can,
for an isotropic background, be captured in a single correction
function
\begin{equation} \label{alphav}
 \alpha_{v,K}=  \frac{1}{3} \sum_{I}
C_{v,I}^{(1/2)}(\{\mu_{v',I'}\})^2
\cdot
 \frac{\sqrt{2\pi\gamma\ell_{\rm P}^2}
 (\mu_{v,I}+\mu_{v,-I})^2}{\prod_{J=1}^3\sqrt{|\mu_{v,J}+\mu_{v,-J}|}}
\end{equation}
which would equal one in the absence of quantum corrections. This is
indeed approached in the limit where all $\mu_{v,I}\gg 1$, but for any
finite values there are corrections. If all $\mu_{v,I}>1$ one can
directly check that corrections are positive, i.e.\ $\alpha_{v,K}>1$
in this regime. Expressing the labels in terms of the densitized
triad through fluxes (\ref{FluxVert}) results in functionals
\begin{equation} \label{alpha}
 \alpha[p^I(v)]= \alpha_{v,K}(4\pi\gamma\ell_{\rm P}^2 \mu_{v,I})
\end{equation}
which enter quantum corrections.

\end{appendix}

\end{document}